\documentclass[iop]{emulateapj}
\usepackage{graphicx}
\usepackage{amssymb}
\usepackage{rotating}
\usepackage{amsmath,wasysym}
\usepackage{natbib}
\begin{document}

\title{Consistent scaling laws in anelastic spherical shell dynamos}

\author{Rakesh K. Yadav\altaffilmark{1}, Thomas Gastine, Ulrich R. Christensen, L\'ucia D. V. Duarte\altaffilmark{2}}
\affil{Max-Planck-Institut f\"ur Sonnensystemforschung, Max Planck Strasse 2, \\ 37191 Katlenburg-Lindau, Germany}

\altaffiltext{1}{Also at the Institut f\"ur Astrophysik, Georg-August-Universit\"at, Friedrich-Hund-Platz 1, 37077 G\"ottingen, Germany}
\altaffiltext{2}{Also at the Technische Universit\"at Braunschweig, Germany}

\email{yadav@mps.mpg.de}

\begin{abstract}
Numerical dynamo models always employ parameter values that differ by orders of magnitude from the values expected in natural objects. However, such models have been successful in qualitatively reproducing properties of planetary and stellar dynamos. This qualitative agreement fuels the idea that both numerical models and astrophysical objects may operate in the same asymptotic regime of dynamics. This can be tested by exploring the scaling behavior of the models. For convection-driven incompressible spherical shell dynamos with constant material properties, scaling laws had been established previously that relate flow velocity and magnetic field strength to the available power.  Here we analyze 273 direct numerical simulations using the anelastic approximation, involving also cases with radius-dependent magnetic, thermal and viscous diffusivities.  These better represent conditions in gas giant planets and low-mass stars compared to Boussinesq models. Our study provides strong support for the hypothesis that both mean velocity and mean magnetic field strength scale as a function of power generated by buoyancy forces in the same way for a wide range of conditions. 
\end{abstract}


\section{Introduction}

Dynamo simulations aim to capture the magnetic field generation process in planetary and stellar interiors and have been very successful in qualitatively reproducing many of the observed properties~\citep{Brun2004, Wicht2010, Jones2011a}. However, all such numerical simulations use inconsistent control parameters, either too large or too small, due to limited computational resources. As a consistency check between our theoretical understanding of the dynamo mechanism with astrophysical observations, it is of paramount importance to establish generic scaling laws which are valid in the relevant  dynamical regimes.

Many attempts have been made earlier to derive a scaling theory for the mean magnetic field, for example, on force balance considerations~\citep{Stevenson1979, Curtis1986, Mizutani1992, Sano1993, Starchenko2002}. However, none of the suggested scaling laws was generally accepted. Progress was  made in this regard due to the increase in computational power in the last decade. \citet{Christensen2006} analyzed the results of parameter studies of dynamo simulations under the Boussinesq approximation in a rotating spherical shell and found that mean velocity and mean magnetic field scale as a function of the available convective power generated via buoyancy forces. The power-based scaling laws (hereafter referred to as PBS) do not explicitly depend on rotation rate in case of magnetic field scaling. In the parameter range of the simulations, a secondary influence  of diffusivities in the form of magnetic Prandtl number (ratio of viscosity and magnetic diffusivity) was found.

Subsequent studies generalized the PBS to Boussinesq dynamos with different physical setups and different boundary conditions~\citep{Takahashi2008, Aubert2009, Christensen2010, Schrinner2012, Yadav2013}. \citet{Olson2006} derived PBS for the magnetic dipole moment from numerical simulations and found an order of magnitude agreement with dynamos in solar system planets. Furthermore, \citet{Christensen2009} and \citet{Christensen2010} observed a good agreement between PBS of magnetic field from numerical simulations and observationally constrained magnetic field of Earth, Jupiter, rapidly rotating low-mass stars, and possibly Uranus and Neptune. \citet{Stelzer2013} maintain that including a dependence on the magnetic Prandtl number is mandatory for an adequate fit, at least in the parameter range of current numerical simulations. \citet{Davidson2013} recently provided interesting theoretical arguments supporting the power based scaling laws.

The agreement between the prediction of scaling laws derived from Boussinesq numerical simulations and observed magnetic fields in low-mass stars is rather puzzling. Unlike the dynamo mechanism in solar type stars, where the strong differential rotation and magnetic field generation by shear at the tachocline are thought to be a key ingredient~\citep{Ossendrijver2003}, rapidly rotating low-mass stars ($\text{mass}<0.35\,M_{\astrosun}$) and giant planets possibly harbor dynamos similar to the geodynamo, where helical convection columns aligned with the rotation axis are instrumental. However, the hydrogen rich interiors of low-mass stars are vastly different from the liquid metal interiors of the Earth-like objects. The density and transport properties in the liquid core of the latter vary by some tens of percent (e.g.~$\approx$ 20\% density change across Earth's liquid core; see ~\citet{Braginsky1995}) and can be considered constant. On the other hand, the interiors of gas planets and stars have significant density stratification and transport properties (such as electrical conductivity and thermal diffusivity) may vary by orders of magnitude~\citep{French2012}.

The power based scaling laws discussed above were derived from a large number of Boussinesq dynamo simulations. Although density stratified models with radially-varying transport properties have been commonly employed in the stellar dynamo community (see e.g.~\citet{Gilman1981, Brun2004, Browning2008}), a systematic scaling study of important diagnostic quantities has never been carried out. One of the reasons is the substantial increase in computational requirements associated with anelastic density-stratified dynamo simulations which makes parameter studies rather expensive~\citep{Jones2009b}. For nonmagnetic rotating convection, \citet{Gastine2012} found in systematic model studies that the velocity of convection and of zonal flow scales in the same way for Boussinesq and anelastic cases. Here we extend this to dynamo models with density stratification. We also include cases with different forms of variation in transport properties, different radial gravity profiles and different mechanical and magnetic boundary conditions. Our aim is not to model any particular class of astrophysical object as realistically as possible, but rather concentrate on generic scaling properties. Our analysis of more than 270 numerical dynamo models shows that the same power based scaling laws apply to a wide variety of dynamos.

\section{Equations and numerical setups}
\subsection{Anelastic MHD equations}
We consider dynamo action in spherical shells, with inner radius $r_i$ and outer radius $r_o$, filled with an electrically conducting fluid. The aspect ratio $\eta$ is defined as $r_{i}/r_{o}$. The shell rotates along a vertical axis $\hat{z}$ with constant angular velocity $\Omega$.  Convection inside the shell is driven by a fixed entropy contrast $\Delta s$ between the inner and the outer boundary. We work in dimensionless units using shell thickness $D=r_{o}-r_{i}$ and inverse rotation frequency $\Omega^{-1}$ as the fundamental length and time units, respectively. The density $\rho$ and entropy $s$ are non-dimensionalized using $\rho(r_o)=\rho_o$ and $\Delta s$, respectively. Magnetic field is scaled by $\Omega D\sqrt{\mu\rho_{o}}$, where $\mu$ is the magnetic permeability.

To model the low-Mach number flows in the density stratified interiors of giant planets and low-mass stars, we employ the anelastic approximation. It allows radial variation of density while filtering out the fast acoustic waves out of the system~\citep{Braginsky1995, Lantz1999}. The anelastic approximation assumes an adiabatic reference state, i.e. $d\tilde{T}/dr=-g/c_{p}$, where $g$ is gravity and $c_{p}$ is the specific heat at constant pressure. The radius-dependent reference state quantities are highlighted with a tilde on top.  For the sake of generality we define the gravity profile as 
\begin{equation}
g(r) = g_{1}\frac{r}{r_{o}} + g_{2}\frac{r^{2}_{o}}{r^{2}}, \label{eq:gravity}
\end{equation}
and by using $g_{1}$ or $g_{2}$ appropriately we can either choose a linear gravity profile ($g_1 = 1$, $g_2 = 0$), approximately representing a self-gravitating body with weak density variation, or an $r^{-2}$ gravity profile ($g_1 = 0$, $g_2 = 1$), exemplifying objects with massive core. Assuming an ideal gas equation of state leads to a polytropic reference state defined by $\tilde{\rho}=\tilde{T}^{m}$, where $m$ is the polytropic index. Solving for $\tilde{T}$ using Eq.~(\ref{eq:gravity}) leads to 
\begin{equation}
\tilde{T}=1-c_{o}\left[\frac{g_{1}}{2}\left(\frac{r^{2}}{r_{o}^{2}}-1\right)+g_{2}\left(1-\frac{r_{o}}{r}\right)\right]
\end{equation}
with 
\begin{equation}
c_{o} = \frac{ \eta\left( e^{\frac{N_{\rho}}{m}} - 1 \right)}{ \frac{g_{1}}{2}(\eta - \eta^{3}) + g_{2}(1-\eta )}, \nonumber
\end{equation}
where $N_{\rho}=\ln(\tilde{\rho}(r_{i})/\tilde{\rho}(r_{o}))$ represents the number of density scale heights across the shell.

The thermodynamic variables, density, pressure, and temperature are then decomposed into the sum of reference state values and small perturbations as $\tilde{\rho}+\rho$, $\tilde{P}+p$, and $\tilde{T}+T$ respectively~\citep{Gilman1981, Braginsky1995, Lantz1999}. The evolution of velocity $\mathbf{u}$ is governed by 
\begin{gather}
\nabla\cdot(\tilde{\rho}{\bf u})=0, \label{eq:div_v} \\
\frac{\partial{\bf u}}{\partial t}+{\bf u}\cdot\nabla\mathbf{u}+2\,\hat{z}\times{\bf u}= -\nabla\frac{p}{\tilde{\rho}}+\frac{Ra\, E^{2}}{Pr}\, g(r)\, s\,\hat{r} \nonumber \\
+\,\frac{1}{\tilde{\rho}}(\nabla\times{\bf B})\times{\bf B}+\frac{E}{\tilde{\rho}}\nabla\cdot \tilde{\nu}S, \label{eq:vel}   
\end{gather}
where $\mathbf{B}$ is magnetic field and $\hat{r}$ is the radial unit vector. The traceless rate-of-strain tensor $S$ is defined by 
\begin{gather}
S_{ij}=2\tilde{\rho}\left(e_{ij}-\frac{1}{3}\delta_{ij}\nabla\cdot\mathbf{u}\right) 
\end{gather}
with 
\begin{gather}
e_{ij}=\frac{1}{2}\left(\frac{\partial u_{i}}{\partial x_{j}}+\frac{\partial u_{j}}{\partial x_{i}}\right), 
\end{gather}
where $\delta_{ij}$ is the identity matrix. The entropy $s$  evolves according to  
\begin{gather}
\tilde{\rho}\tilde{T}E\left(\frac{\partial s}{\partial t}+{\bf u}\cdot\nabla s\right)=\frac{E^{2}}{P_{r}}\nabla\cdot(\tilde{\kappa}\tilde{\rho}\tilde{T}\nabla s) \nonumber \\
+\frac{P_r\,c_{o}\,(1-\eta)}{Ra}\left[\tilde{\nu}\,Q_{\nu}+\frac{\tilde{\lambda}}{P_{m}}(\nabla\times\mathbf{B})^{2}\right], \label{eq:entropy}
\end{gather}
where the viscous heating contribution is 
\begin{gather}
Q_{\nu}=2\tilde{\rho}\left[ e_{ij} e_{ji} - \frac{1}{3} (\nabla\cdot\mathbf{u})^2 \right].
\end{gather}
The magnetic induction is governed by
\begin{gather}
\frac{\partial{\bf B}}{\partial t}=\nabla\times({\bf u}\times{\bf B})-\frac{E}{P_m}\nabla\times\left(\tilde{\lambda}\nabla\times{\bf B}\right), \label{eq:mag} \\  
\nabla\cdot{\bf B}=0. \label{eq:div_B} 
\end{gather}

In the above equations, kinematic viscosity $\nu$, thermal diffusivity $\kappa$, and magnetic diffusivity $\lambda$ are normalized by their value at the inner boundary ($r=r_i$); they are denoted by $\tilde{\nu}$, $\tilde{\kappa}$, and $\tilde{\lambda}$. The various dimensionless control parameters appearing in Eqs.~(\ref{eq:vel}-\ref{eq:mag}) are:
\begin{gather}
\text{Rayleigh number } Ra=\frac{g(r_{o})D^{3}\Delta s}{c_{p}\,\nu_{i}\,\kappa_{i}}, \nonumber \\ 
\text{ Ekman number } E=\frac{\nu_{i}}{\Omega D^{2}}, \nonumber  \\ 
\text{Prandtl number } P_r=\frac{\nu_{i}}{\kappa_{i}}, \nonumber \\
\text{ magnetic Prandtl number }P_m=\frac{\nu_{i}}{\lambda_{i}}, \nonumber 
\end{gather}
where the $``i"$ subscript represents values at the inner boundary. 

\subsection{\label{sec:var_prop}Variable properties}
Recent {\it ab initio} calculations suggest that electrical conductivity decreases with radius by several orders of magnitude in the outer regions of Jupiter-like objects~\citep{French2012}. Low-mass stars, brown dwarfs and massive extrasolar planets will probably show a similar variation in electrical conductivity. An electrical conductivity profile approximately constant in the deep  interior and exponentially decaying in outer portions represents a good model for giant planets~\citep{French2012}.
We model electrical conductivity normalized by its value at the inner boundary   as~\citep{Gomez2010, Duarte2013} 
\begin{gather}
\tilde{\sigma}(r)=
\begin{cases}
1+(\tilde{\sigma}_{m}-1)\left(\frac{r-r_{i}}{r_{m}-r_{i}}\right)^{a} & r<r_{m}, \\
\tilde{\sigma}_m e^{a\left( \frac{r-r_m}{r_m-r_i}\right)\frac{\tilde{\sigma}_m-1}{\tilde{\sigma}_m}} & r \ge r_m
\end{cases}. \label{eq:Lucia}
\end{gather}
The electrical conductivity follows a polynomial in $r_i<r<r_m$ and decreases to $\tilde{\sigma}(r_m)=\tilde{\sigma}_m$ (usually 0.5) near $r_m$. The exponential decay starts for $r\ge r_m$. The constant $a$ defines the rate of decay; $a$ is mostly equal to 9, except of two cases where it is 1 and 25. The relative transition radius $\chi_m=r_m/r_i$ is 0.7, 0.8, 0.9, or 0.95. The magnetic diffusivity $\lambda=(\mu\,\tilde{\sigma}(r))^{-1}$ accordingly rises along the radius. We use such profiles in many of our anelastic dynamo simulations.

We also simulate cases with diffusivities, viz. viscosity, thermal diffusivity, magnetic diffusivity, varying as a function of density: 
\begin{gather}
(\nu,\kappa,\lambda)=(\nu,\kappa,\lambda)_{i}\sqrt{\frac{\tilde{\rho}_{i}}{\tilde{\rho}(r)}}. \label{eq:ASH}
\end{gather}
Such density dependence is similar to that used in many stellar dynamo simulations (e.g.~\citet{Brun2004, Browning2008}).

\subsection{Boundary conditions}
The mechanical boundary condition is either stress-free on both boundaries or rigid on the inner and stress-free on the outer boundary. On both boundaries constant entropy is imposed.  We do not simulate cases with flux boundary conditions. For Boussinesq cases, no difference in scaling laws had been found between fixed temperature and fixed flux conditions~\citep{Aubert2009, Christensen2010}. The inner core is either conducting or insulating. The magnetic field matches a diffusive solution at the inner boundary in case of a conducting inner core and a potential field in case of an insulating inner core, while it always matches a potential field on the outer boundary.

\subsection{Numerical technique}
The anelastic equations (Eqs.~\ref{eq:div_v}-\ref{eq:div_B}) are time advanced using MagIC~\citep{Wicht2002, Gastine2012}. The anelastic version of MagIC has been benchmarked with recent community-based benchmark simulations~\citep{Jones2011b}. The mass-flux and the magnetic field are decomposed into poloidal an toroidal parts as
\begin{gather}
\tilde{\rho}\mathbf{u}=\nabla\times(\nabla\times W\hat{r}) + \nabla\times X\hat{r}, \nonumber \\ 
\mathbf{B}=\nabla\times(\nabla\times Y\hat{r}) + \nabla\times Z\hat{r} \nonumber
\end{gather}
where $W$ and $X$ ($Y$ and $Z$) are poloidal and toroidal scalar potentials for mass-flux (magnetic field). The scalar potentials, along with $p$ and $s$, are then expanded using spherical harmonic functions in the azimuthal and the latitudinal direction. The radial expansion is done using Chebyshev polynomials. The equations are time-stepped by advancing non-linear and Coriolis terms using an explicit second-order Adams-Bashforth scheme while the remaining terms are time-advanced using an implicit Crank-Nicolson algorithm (see \citet{Glatzmaier1984} and \citet{Christensen2007} for more details).

\section{Results}
\subsection{Diagnostic parameters}
In the following discussions we will use several non-dimensional diagnostic parameters which are derived from the numerical simulations. They describe representative mean values of the flow velocity (or kinetic energy) and magnetic field strength (or magnetic energy). While in the Boussinesq case these properties have similar amplitude throughout the shell, with density stratification their magnitude can vary significantly with radius, in particular for the velocity. Hence the question arises which property to average in which way. We found that averaging energies divided by unit mass gives the best results.

The non-dimensional rms velocity is given by the Rossby number 
\begin{gather}
Ro=\sqrt{\frac{1}{V}\int{(\mathbf{u}\cdot\mathbf{u})}\,dV} \label{eq:Ro_def}
\end{gather}
where $V$ is volume of the fluid shell and the volume integral is evaluated in the fluid shell. The non-dimensional kinetic energy per unit mass is then $Ro^2/2$. The Lorentz number $Lo$ describes the non-dimensional magnetic field strength and is defined as
\begin{gather}
Lo=\sqrt{\frac{\int{(\mathbf{B}\cdot\mathbf{B})}\,dV}{\int{\tilde{\rho}\,dV}}} \label{eq:Lo_def}
\end{gather}
with $E_m=Lo^2/2$ being the non-dimensional magnetic energy per unit mass. The non-dimensional power generated per unit mass by thermal buoyancy, scaled by $\Omega^3 D^2$, is 
\begin{gather}
P=\frac{Ra\, E^{2}}{P_r}\frac{\int{u_r\,\tilde{\rho}\,g\,s}\,dV}{\int{\tilde{\rho}\,dV}}, \label{eq:p_def}
\end{gather}
where $u_r$ is the radial component of $\mathbf{u}$. The Nusselt number $Nu$ is the ratio of total transported heat flux to the conducted heat flux.  The rate of energy dissipated per unit mass by ohmic dissipation is
\begin{gather}
D_\lambda=\frac{E}{P_m}\frac{\int\tilde{\lambda}\,(\nabla\times\mathbf{B})^2\, dV}{\int{\tilde{\rho}\,dV}}. \label{eq:ohm_def}
\end{gather}
The ohmic fraction 
\begin{gather}
f_{ohm}=D_\lambda /P
\end{gather}
is the fraction of energy dissipated by joule dissipation alone. The characteristic time scale of magnetic energy dissipation is
\begin{gather}
\tau_{mag}=\frac{E_{m}}{D_\lambda}. 
\end{gather}
All diagnostic quantities are time-averaged after a statistically stationary state has been reached in the simulations. It would have been more consistent if in Eqs.~(\ref{eq:Lo_def}-\ref{eq:ohm_def}) the magnetic energy, power, and dissipation would have been mass-normalized before integration (as is done in the case of the kinetic energy) instead of normalizing the total energy by the integral of $\tilde{\rho}$ (which is the non-dimensional total mass). However, the integrations have been done during the simulation and could not be repeated in modified form without repeating the whole run. Because the magnitude of these properties seems to vary less with radius than that of velocity, the difference is probably not critical.

We separate the dynamos resulting from our numerical simulation in two categories: the dipolar category contains dynamos with a dominant axisymmetric magnetic dipole (spherical harmonic degree $\ell=1$ and harmonic order $m=0$); the multipolar class contains all other kinds of dynamos. To carry out this separation we use the dipolarity $f_{dip}$ which is the magnetic energy in the axisymmetric dipole normalized by the cumulated magnetic energy in  harmonic degrees up to 12, both evaluated at the outer boundary. Dynamos with $f_{dip}>0.3$ are considered dipolar. It must be noted that our data spans the range $10^{-4}<f_{dip}<0.9$ quite uniformly, hence a cutoff of 0.3 remains somewhat arbitrary. We justify our choice of cutoff in Sec.~\ref{sec:mag}.

\subsection{Dynamo database}
Important physical attributes of various numerical setups used in this study are tabulated in Tab.~\ref{tab:tab1}. The database incorporates many important features, such as density stratification and variable transport properties. The largest density contrast in our simulation is $N_{\rho}=5.5$, i.e. $\rho_{i}/\rho_{o}\approx245$. The aspect ratio $\eta$ is varied to cover dynamos operating in thick shells or nearly full spheres as well as dynamos in thinner shells (from $\eta=0.1$ to $\eta =0.75$). The polytropic index $m$ is also changed in a few cases to model different polytropic states. With these features our simulations seek to represent, in a simplified way, the dynamo action in the fully or partially convective and density stratified interiors in different classes of objects ranging from giant planets to low-mass stars.
\begin{table*}
\footnotesize
\begin{center}
\caption{Simulation setups} \label{tab:tab1} 
\begin{tabular}{ccccccccccc}
\hline \hline
\parbox[t]{3cm}{$\,\,$Subset name \\(number of cases)} & Core &BC & $\eta$ & $g(r)$ & $m$ & $Pr$ & $Pm$ & $E$ & $N_{\rho}$ & \parbox[t]{2.5cm}{Varying inside\\ the shell}\tabularnewline
\hline 
Anelastic 	&  				&  				&  		&  			&  	  & & & & 							& \tabularnewline
A1 (8) 		& Insulating 	&$\text{SF}$ 	& 0.1 	& r 		& 2   & 1 -- 10 & 0.5 -- 10 & $3 \times 10^{-6}$ -- $1 \times 10^{-4} $ & 2, 4  	 & none\tabularnewline
A2a (44) 	& Conducting 	&$\text{mixed}$ & 0.2 	& r 		& 2   & 1 		& 1 -- 10   & $1 \times 10^{-5}$ -- $1 \times 10^{-3} $ & 0.5 -- 5  &  none\tabularnewline
A2b (43) 	& Conducting 	& mixed 		& 0.2 	& r 		& 2   & 1		& 0.5 -- 5  & $1 \times 10^{-5}$ -- $3 \times 10^{-4} $ & 1 -- 5.5   &$\lambda$ \tabularnewline
A3a (58) 	&Insulating 	& SF 			& 0.35 	& r 		& 2   & 0.3 -- 10& 0.4 -- 20 & $1 \times 10^{-5}$ -- $1 \times 10^{-3} $ & 0.5 -- 5  				&  none\tabularnewline
A3b (9) 	&Insulating 	& SF 			& 0.35 	& 1/$r^{2}$ 	& 1.5 & 0.5, 1& 0.5 -- 10 & $3 \times 10^{-5}$ -- $1 \times 10^{-3} $& 1, 3   	  				&  none\tabularnewline
A4 (11) 	&Insulating 	& SF 			& 0.35 	& 1/$r^{2}$ 	& 2   & 1, 3  & 0.2 -- 3 & $1 \times 10^{-6}$ -- $1 \times 10^{-4} $ & 3, 5  	  				&  $\nu$, $\kappa$, $\lambda$ \tabularnewline
A5 (24) 	&Insulating 	& SF 			& 0.6  	& 1/$r^{2}$ 	& 2   & 1	  & 2        & $1 \times 10^{-4}$ & 0.01 -- 3 				&  none\tabularnewline
A6 (1) 		&Insulating 	& SF 			& 0.75 	& 1/$r^{2}$ 	& 2   & 0.5   & 0.5      & $3 \times 10^{-5}$& 1  		  				&  none\tabularnewline
\hline
Boussinesq &  &  &  &  &  &  & \tabularnewline
B1a (14) & Conducting & mixed & 0.2  & r & 0 & 1 & 0.5 -- 10 & $1 \times 10^{-5}$ -- $1 \times 10^{-3} $ & 0 &  none\tabularnewline
B1b (21) & Conducting & mixed & 0.2  & r & 0 & 1 & 1 -- 10 & $1 \times 10^{-5}$ -- $3 \times 10^{-4} $& 0 &  $\lambda$ \tabularnewline
B2 (40)  &Insulating  & SF 	  & 0.35 & r & 0 & 1 & 0.2 -- 10& $1 \times 10^{-5}$ -- $1 \times 10^{-3} $& 0 & none\tabularnewline
\hline 
\end{tabular}
\end{center}
Notes: BC -- mechanical boundary conditions, where SF stands for shear stress free on both boundaries and mixed stands for a rigid inner and stress-free outer boundary, aspect ratio $\eta$, gravity profile $g(r)$, polytropic index $m$, Prandtl number $Pr$, magnetic Prandtl number $Pm$, Ekman number $E$, and number of density scale heights inside the shell $N_{\rho}$ for various simulation subsets. The last  columns list transport properties which vary along the shell radius. In groups A2b and B1b the variation of $\lambda$ is obtained from Eq.~\ref{eq:Lucia} and in group A4 the variation follows Eq.~\ref{eq:ASH}.\\
\end{table*}

In total, 273 dynamo cases are simulated. To the best of our knowledge this is the largest database of this kind. A portion of this database has already been used in earlier studies to explore different aspects of dynamo mechanism: A2a, B1a, A5 from \citet{Gastine2012dyn} and \citet{ Gastine2013a}; A2b, B1b from \citet{Duarte2013}; A3a from \citet{Yadav2013}.  The full database can be found in the online supplementary table. More details, e.g.~on the choice of model parameters, can be found in the references mentioned before. The simulations cover a large parameter space: $0.3 \le P_r \le 10$, $0.2 \le P_m \le 20$, $1\times 10^{-6}\le E \le 1\times 10^{-3}$, and $2.5\times 10^{5} \le Ra \le 2.5\times 10^{9}$. Except for two lowest Ekman number simulations, all simulations were run until the simulated time was at least one magnetic diffusion time ($D^{2}/\lambda$) and a statistically stationary temporal behavior was acquired. Depending on $Pr$ and $Pm$ viscous ($D^{2}/\nu$) and thermal ($D^{2}/\kappa$) diffusion time scales could be  larger than the magnetic diffusion time, but because the magnetic field tends to equilibrate more slowly than the thermal or the velocity field, the latter time scale is more relevant. The range of our data-set is dictated by the computational need of a simulation. For example, low $E$ simulations are computationally demanding, but such dynamos can be obtained at lower $Pm$ and hence as $E$ decreases so do our explored $Pm$ values. No azimuthal symmetry was imposed in the simulations, except for the large aspect ratio run of A6 where two fold symmetry in longitude was used.

Our aim is to explore scaling laws in the rotationally dominated dynamic regime, i.e.~Coriolis forces are dominant and inertia plays a secondary role, hence we report and analyze dynamos with $1\times 10^{-6}\le E \le 1\times 10^{-3}$ and avoid convection with very high supercritical Rayleigh numbers. Furthermore, to ensure fully developed convection in the spherical shell we use only dynamos with $Nu\ge2$.

\subsection{Velocity power scaling}
Scaling laws for velocity and magnetic field in Boussinesq models have been conventionally expressed in terms of flux-based Rayleigh number $Ra_{Q}^{*}=Ra\,(Nu-1) E^{3}/Pr$ which serves as a proxy for non-dimensional power $P$~\citep{Christensen2002, Christensen2006, Olson2006, Takahashi2008, Schrinner2012, Yadav2013, Schrinner2013}. $Ra_{Q}^{*}$ is in general directly proportional to $P$ but the proportionality constant is determined by the particular simulations setup. In Boussinesq models the proportionality constant can be calculated in terms of shell aspect ratio and the distribution of buoynacy sources and sinks~\citep{Christensen2006, Aubert2009}, but for our anelastic models its dependence on density stratification and variable transport coefficients becomes nontrivial. Following \citet{Kaspi2009}, \citet{Gastine2012} expressed the velocity scaling in their anelastic convection simulations using a mass averaged ${\langle Ra_{Q}^{*}\rangle}_{\rho}$, where $\langle \cdot \rangle_{\rho}$ stands for $\int (\cdot)\,\tilde{\rho}\,r^{2}\,dr/\int\,\tilde{\rho}\,r^{2}\,dr$, to better collapse simulations with different density stratifications. However, different scaling laws in our database, which includes cases with radially varying transport coefficients, do not collapse well with ${\langle Ra_{Q}^{*}\rangle}_{\rho}$ as a control parameter. We find that the power $P$, whose time-average value is calculated explicitly in each simulation, is a much better control parameter for collapsing diverse simulation setups.

\begin{figure}
\epsscale{1.15}
\vspace{7mm}
\plotone{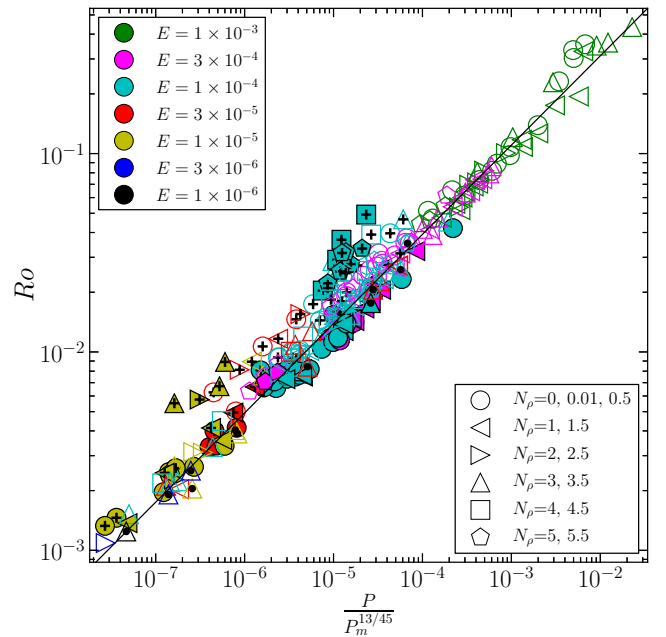}
\caption{Rossby number versus a combination of non-dimensional power and magnetic Prandtl number. The solid line represents $Ro=2.47\,P^{0.45}\,{P_m}^{-0.13}$. Filled (empty) symbols display dipolar (multipolar) dynamos. The Ekman number is color coded and the symbol shape indicates degree of density stratification. Symbols containing a ``+" marker have exponentially decaying electrical conductivity in the outer regions (subset A2b and B1b in Tab.~\ref{tab:tab1}) and the ones carrying a dot symbol have moderate conductivity variations (subset A4 in Tab.~\ref{tab:tab1}). \label{Ro}}
\end{figure}

Least-square-optimization of the Rossby number $Ro$ as a function of power $P$ and magnetic Prandtl number $P_m$ for our database results in
\begin{gather}
Ro=2.47\,\frac{P^{0.45}}{{P_m}^{0.13}}. \label{eq:Ro}
\end{gather}
This empirical scaling is represented by a solid line in Fig.~\ref{Ro}. The dynamo cases marked by a ``+" in Fig.~\ref{Ro} (subset A2b and B1b; see Tab.~\ref{tab:tab1}) are excluded from the fit. The excluded cases have   exponentially decaying electrical conductivity in the outer parts of the spherical shell. In these regions Lorentz forces are weak which allows for the development of strong zonal flows that enhance the rms velocity~\citep{Duarte2013}. This is evident in Fig.~\ref{Ro} where most of these cases have higher $Ro$ and lie above other dynamo cases. The dynamo cases from subset A4 with moderate conductivity variation, which are marked by a black dot in Fig.~\ref{Ro}, do not show enhanced zonal flow and their Rossby numbers agree well with that of the other dynamos.

The scatter of data points in Fig.~\ref{Ro} is primarily due to differences in the relative strength of zonal flows~\citep{Yadav2013}. Magnetic field geometry, which can vary substantially from one simulation to another, affects the zonal flow through Maxwell stresses~\citep{Browning2008}. \citet{Yadav2013} reported two different $Ro$ scalings for dipolar and multipolar dynamos (subset A3a; see Tab.~\ref{tab:tab1}), although the difference was relatively minor. They argued that this offset is due to more efficient zonal flow quenching by magnetic braking in dipolar dynamos as compared to the multipolar ones. Such consistent dichotomy is washed out in Fig.~\ref{Ro}, probably due to our much more diverse dataset.

\begin{figure}
\epsscale{1.15}
\plotone{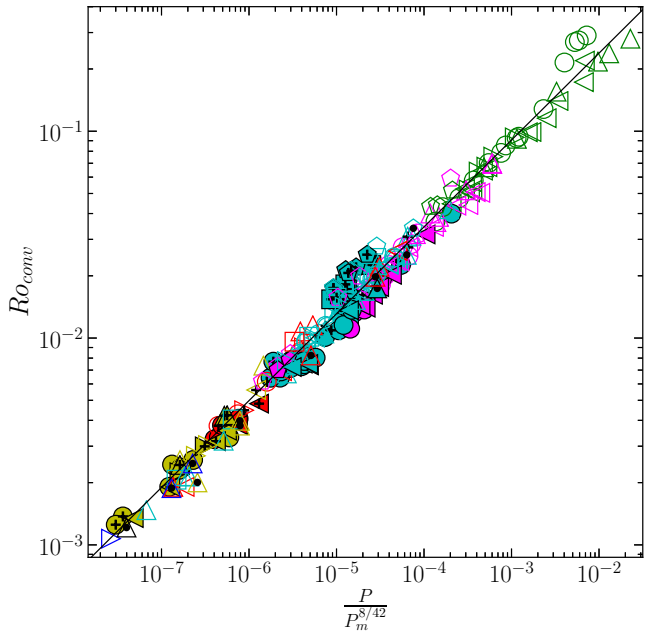}
\caption{Convective Rossby number versus a combination of power and magnetic Prandtl number. The solid line represents $Ro_{conv}=1.65\,P^{0.42}\,{P_m}^{-0.08}$. Refer to Fig.~\ref{Ro} for symbol description. \label{Ro_conv}}
\end{figure}

On similar lines as~\citet{Yadav2013}, we define a convective Rossby number $Ro_{conv}$ which is calculated by excluding  zonal  and  meridional flow components. A best-fit of $Ro_{conv}$ gives 
\begin{gather}
Ro_{conv}=1.65\,\frac{P^{0.42}}{P_m^{0.08}}, \label{eq:Ro_conv}
\end{gather}
which is portrayed in Fig.~\ref{Ro_conv}. As compared to Fig.~\ref{Ro} the scatter in Fig.~\ref{Ro_conv} is much smaller and also cases with a poorly conducting exterior region are now well fitted. On the other hand, the Rossby number based on the zonal flow component is strongly scattered in our database (not shown) and does not exhibit a consistent scaling, in contrast to what has been found in purely hydrodynamic spherical shell convection~\citep{Aubert2001, Christensen2002, Aubert2005, Showman2011, Gastine2012}. As discussed above, this is due to the variable influence of magnetic field on the zonal flow via Maxwell stresses.

A scaling law of the type $Ro\propto {P}^{\alpha}$ with $0.41\le\alpha\le 0.44$ has been reported by many studies of Boussinesq numerical dynamo simulations: in a wide control parameter study with rigid mechanical boundaries~\citep{Christensen2006, Takahashi2008}; for compositional and thermal convection with variable core size, including cases with nearly fully convective interiors~\citep{Aubert2009}; in simulations with different thermal boundary conditions~\citep{Christensen2010}; and for cases with stress-free mechanical boundaries~\citep{Yadav2013}. 

The theoretical justifications for the velocity scaling $Ro \propto P^\alpha$, with $\alpha$ slightly larger than 0.4, is not entirely clear. \citet{Christensen2010} discussed several proposed scaling approaches and showed that they all lead to a law of this form, but with different exponents. The mixing length theory~\citep{Bohm1958} applied to planetary interiors predicts $\alpha=1/3$, but is probably not applicable to rapidly rotating systems~\citep{Christensen2010}. \citet{Aubert2001} considered a triple force balance of Coriolis, inertial, and Archimedean forces (the so-called CIA balance) in the hydrodynamic fluid vorticity equation and derived $\alpha=2/5$. \citet{Starchenko2002} obtained $\alpha=1/2$ based on a balance of Coriolis and buoyancy forces. The prediction of the CIA balance is the closest to the exponent found by fitting the data from dynamo simulations, but underpredicts it slightly. \citet{Davidson2013} presented a scaling theory starting from the assumption that the magnetic field is independent of rotation rate and scales with the cubic root of the power, as confirmed by the numerical results (see next section). \citet{Davidson2013} considers a triple force balance of Lorentz, Archimedean, and Coriolis forces (the so called MAC balance) to then derive a velocity scaling of the form $Ro\propto P^{4/9}$. The exponent is very close to our empirically obtained exponents $\alpha=0.45$ (Eq.~\ref{eq:Ro}) or $\alpha=0.42$ (Eq.~\ref{eq:Ro_conv}). None of the scaling theories for velocity proposed so far accounts for the $P_m$ dependence. Inclusion of $P_m$ as a fit parameter significantly reduces the scatter of data-points~\citep{Christensen2006, Yadav2013, Stelzer2013}.

\subsection{\label{sec:mag}Magnetic field scaling}
\begin{figure}
\epsscale{1.15}
\plotone{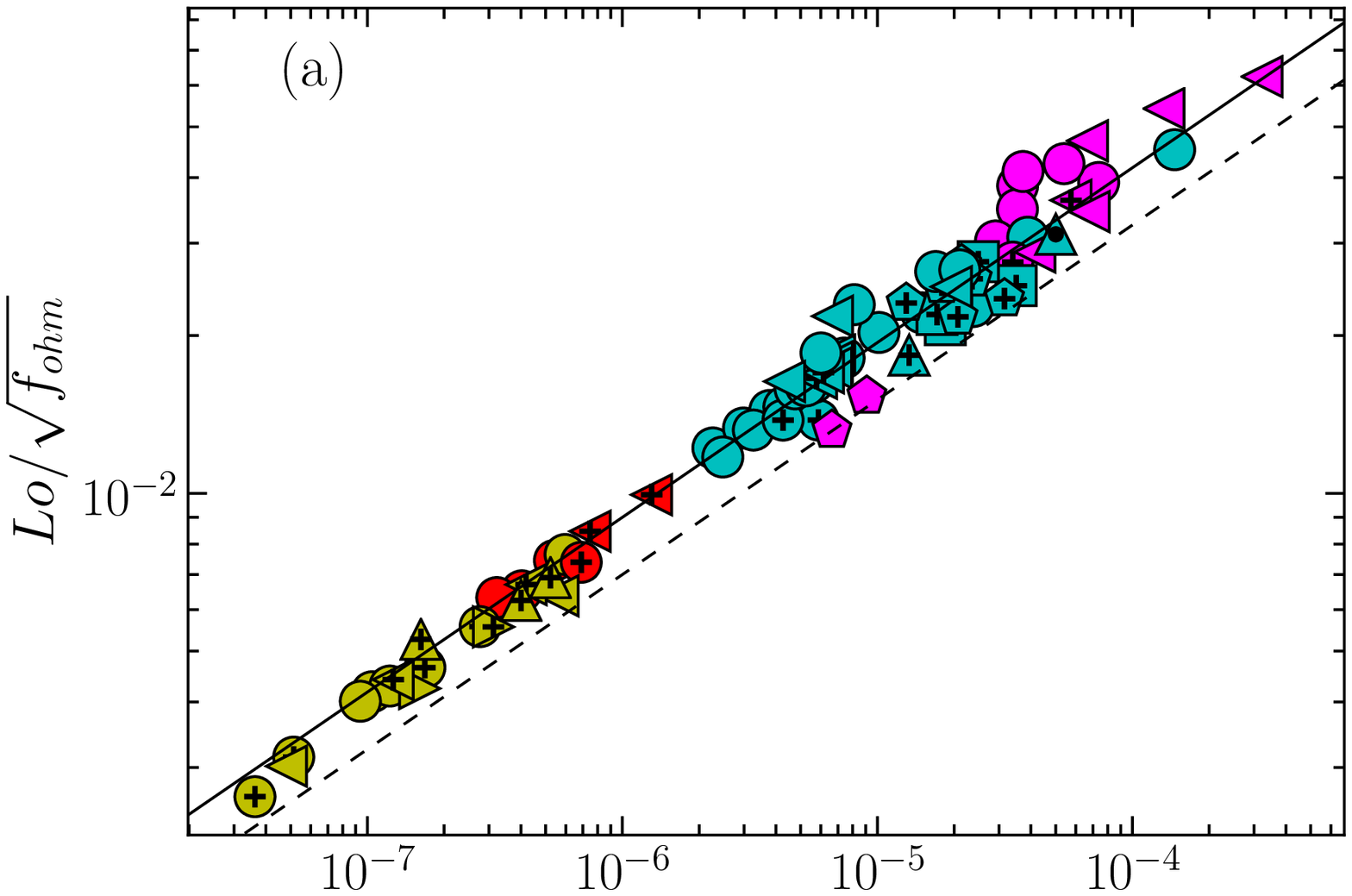}
\vspace{10mm}
\plotone{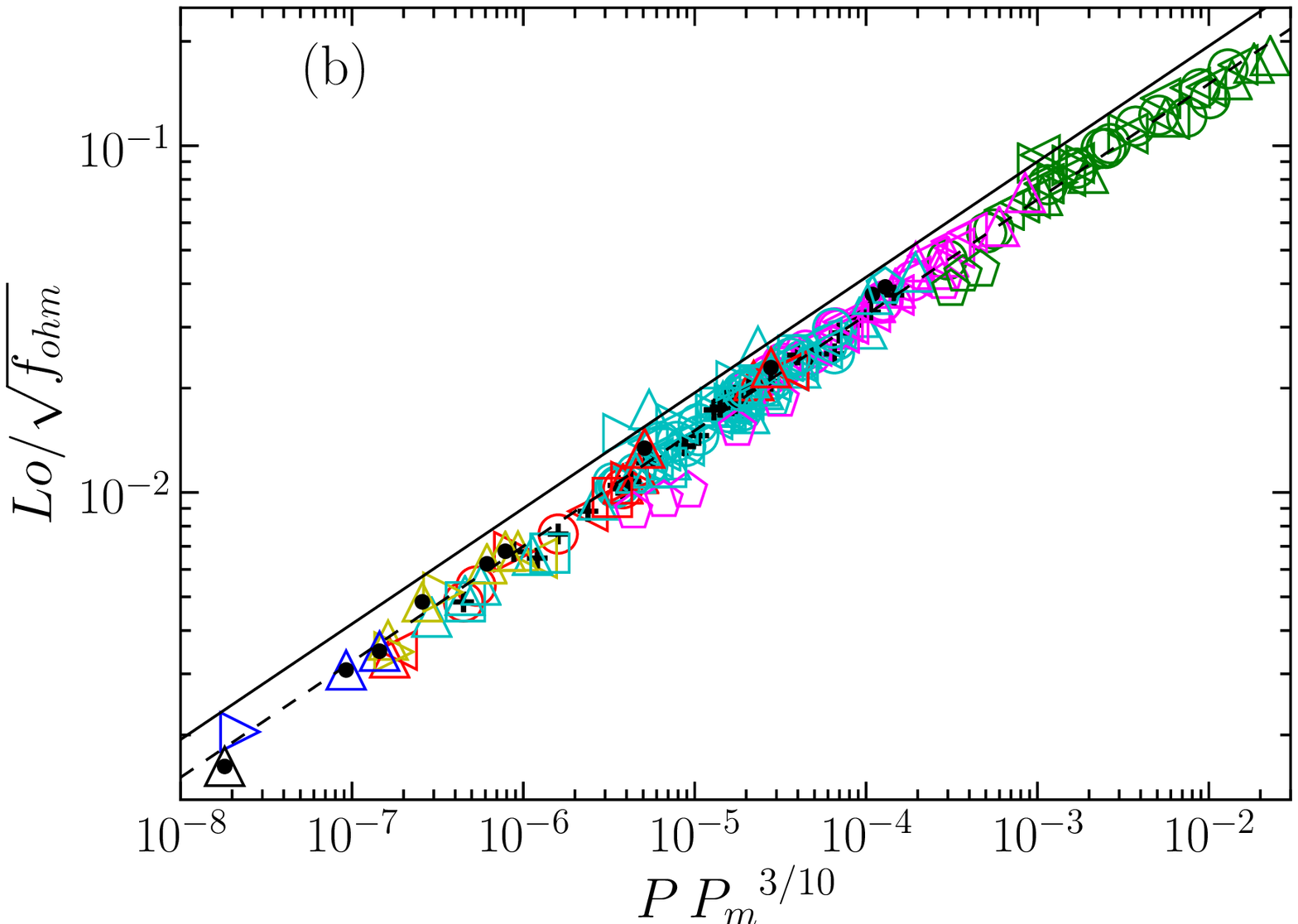}
\caption{Lorentz number corrected for fraction of ohmic dissipation versus a combination of non-dimensional power and magnetic Prandtl number.  (a) dipolar dynamos, the fit by the solid line follows Eq.~(\ref{eq:Lofohm_sim}) with $c=0.9$. (b) multipolar dynamos, the fit by the broken line is for $c=0.7$. The two fitting lines are replicated in both plots to show the offset. Refer to Fig.~\ref{Ro} for symbol description. \label{Lofohm_sim}}
\end{figure}

In the Boussinesq dynamo simulations, \citet{Christensen2010}, \citet{Schrinner2012}, and \citet{Yadav2013} found that the magnetic field strength obeys very similar scaling laws for dipolar and for multipolar dynamos, except for a difference in the pre-exponential constant. Therefore we treat the two classes of dynamos separately. A least-square fit of the Lorentz number corrected by $f_{ohm}$ as a function of $P$ and $P_m$ leads to   
\begin{gather}
\frac{Lo}{\sqrt{f_{ohm}}}=1.08\,P^{0.35}\,{P_m}^{0.13} \label{eq:Lo_dip}
\end{gather}
for the dipolar dynamos and 
\begin{gather}
\frac{Lo}{\sqrt{f_{ohm}}}=0.78\,P^{0.34}\,{P_m}^{0.09} \label{eq:Lo_multi}
\end{gather}
for the multipolar ones. These two shifted scalings provide a basis for separating dynamos in dipolar and multipolar categories. We found that $f_{dip}=0.3$ provides a reasonably good cutoff; lower or higher cutoffs start to mix data points and we loose the demarcation noticeable in Fig.~\ref{Lofohm_sim}.

The two expressions (\ref{eq:Lo_dip}) and (\ref{eq:Lo_multi}) are very similar except for the larger prefactor in the dipolar scaling. We therefore attempt to fit both classes of dynamos with the same approximate exponents for $P$ and $P_m$
\begin{gather}
\frac{Lo}{\sqrt{f_{ohm}}}\approx c\,P^{\frac{1}{3}}\,{P_m}^{\frac{1}{10}} \label{eq:Lofohm_sim}
\end{gather}
where c is fitted separately for dipolar and for multipolar dynamos, giving 0.9 and 0.7, respectively. This approximate scaling is plotted in Fig.~\ref{Lofohm_sim}. \citet{Yadav2013} argued that the higher field strengths in dipolar dynamos can be attributed to lower flow velocity in dipolar dynamos (lack of strong zonal flows). Such argument may not work for our database since both dipolar and multipolar dynamos are consistent with a single velocity scaling (see previous section). Further study is required to explain this behaviour.

Two different theoretical approaches have been put forward to explain the scaling of the Lorentz number with (approximately) the cubic root of the power. \citet{Christensen2006} take a velocity scaling $Ro \propto P^{0.4}$ as given and use the results of \citet{Christensen2004} that the magnetic dissipation time scale $\tau_{mag}$ (in rotational units) is approximately proportional to $Ro^{-1}$. In a different approach, \citet{Davidson2013} uses the finding by \citet{Kunnen2010} that the small scale vorticity in rotating convection is independent of rotation rate. Assuming that this also holds for vorticity at the magnetic dissipation length scale $l_{min}$, for which the local (scale-dependent) magnetic Reynolds number is of order one, he infers that also $l_{min}^2 / \lambda \propto \tau_{mag}$ must be independent of the rotation rate. Setting the $f_{ohm}$ factor aside, the magnetic energy is the product of power and magnetic dissipation time and is therefore also independent of the rotation rate. If magnetic field is only a function of power, dimensional arguments dictate that it must depend on the cubic root of the power.

\begin{figure}
\epsscale{1.15}
\plotone{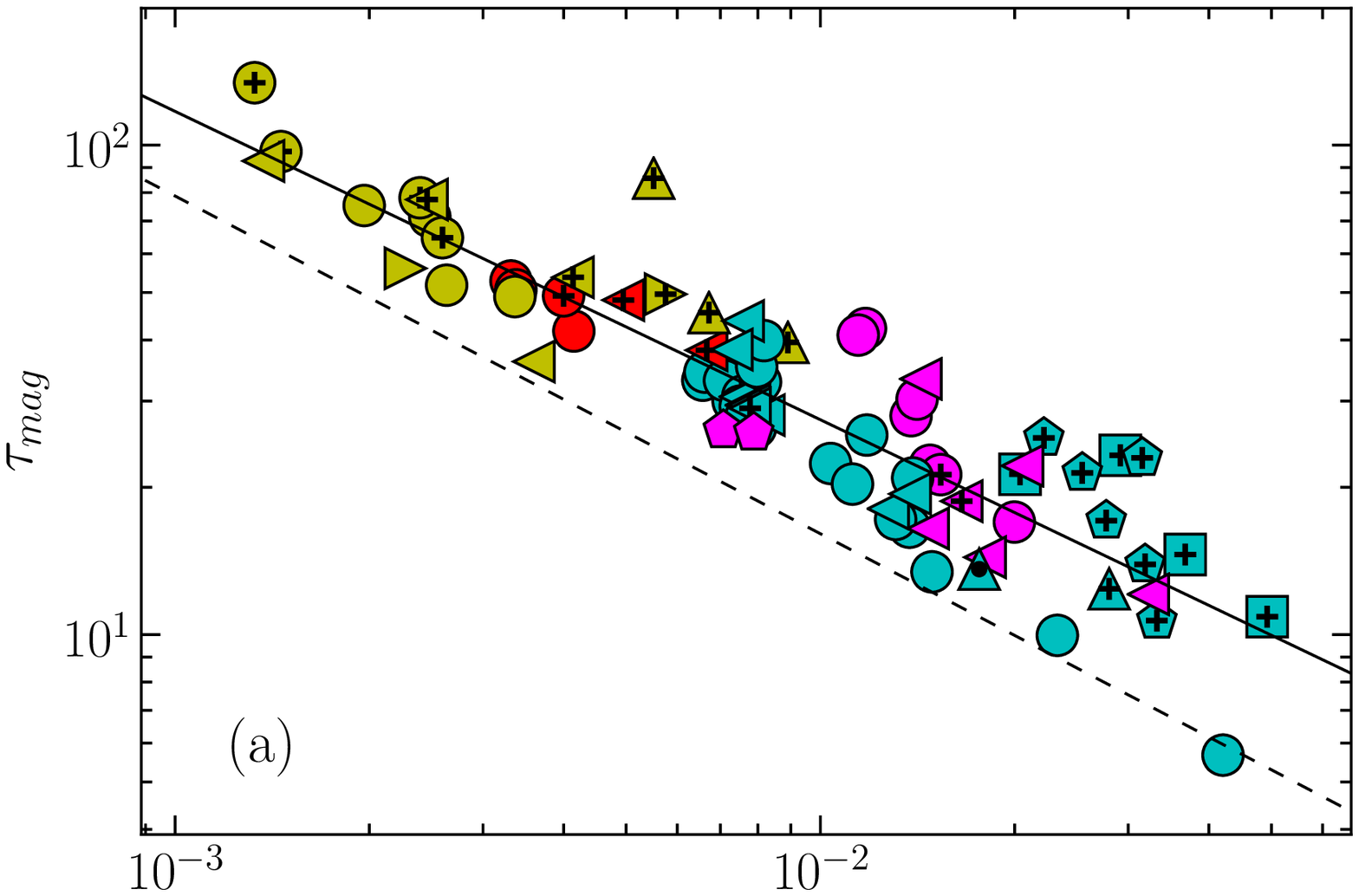}
\vspace{10mm}
\plotone{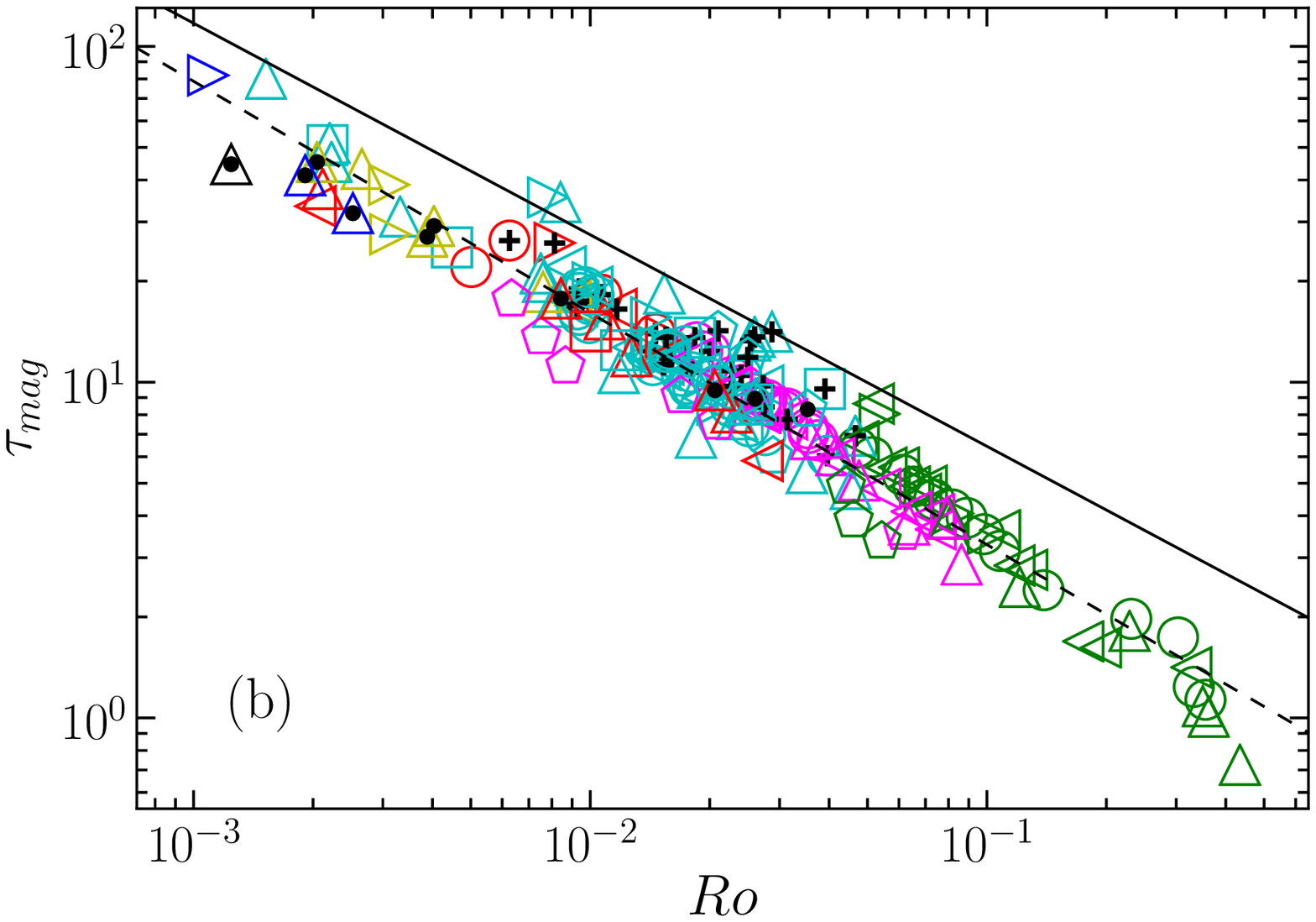}
\caption{Magnetic dissipation time versus Rossby number. (a) dipolar dynamos, (b) multipolar dynamos. The solid and broken lines for the best fits to the two subsets (Eq.~(\ref{eq:tmag_dip}) and (\ref{eq:tmag_multi})) are shown in both panels. Refer to Fig.~\ref{Ro} for symbol description. \label{tmag}}
\end{figure}

Davidson's scaling theory predicts a dependence of the magnetic diffusion time on the Rossby number as $\tau_{mag} \propto Ro^{-3/4}$, significantly weaker than the $Ro^{-1}$ dependence originally envisaged by \citet{Christensen2004}. The scaling of $\tau_{mag}$ vs. $Ro$ in numerical dynamo models had been revisited by \citet{Christensen2010}, \citet{Yadav2013}, and \citet{Stelzer2013} who all found exponents somewhat weaker than -1. In Fig.~\ref{tmag} we plot $\tau_{mag}$ for dipolar and multipolar dynamos against $Ro$. The scatter is larger, especially for dipolar dynamos, than in the cases of fitting $Ro$ and $Lo$. A best-fit line for the dipolar dynamos is 
\begin{gather}
	\tau_{mag} = 1.51\,Ro^{-0.63} \label{eq:tmag_dip}
\end{gather}
and for multipolar dynamos it is
\begin{gather}
	\tau_{mag} = 0.67\,Ro^{-0.69}. \label{eq:tmag_multi}
\end{gather}
This shows that the scaling exponent may be even slighly weaker than the recent prediction by \citet{Davidson2013}.

\section{Discussion and conclusions}
In this study we analyze a large number of numerical simulations of dynamo in rapidly rotating spherical shells to infer empirical scaling laws for mean velocity and mean magnetic field. Our database contains anelastic compressible dynamos with different shell thicknesses, different density stratifications with density contrast upto $\approx 245$, and radially varying transport properties (viscosity, thermal diffusivity, magnetic diffusivity). These models capture, in a simplified way, conditions that are relevant for rapidly rotating gas planets and low-mass stars. 

Our study shows that it is possible to relate flow velocity and magnetic field strength to the available power, by a single scaling law in each case, for a wide range of different conditions. In particular the same scaling law covers incompressible cases and cases with strong density stratification. Our finding sanctifies the application of the same magnetic field scaling by \citet{Christensen2009} to such diverse objects as the Earth's core and low-mass stars.

Our best-fitting power exponent of 0.45 for the Rossby number, as a measure of the characteristic velocity, is very close to the value of 4/9 obtained by the scaling theory of \citet{Davidson2013} which assumes a MAC balance. Our data are better collapsed when the Rossby number is calculated only with the convective flow (excluding the zonal and meridional flow component) and the optimal exponent of 0.42 is intermediate between Davidson's prediction and the value of 2/5 inferred from a CIA balance \citep{Aubert2001}. Because our simulations may not have exactly reached an asymptotic regime, the precise value of the exponent and its theoretical justification remain somewhat uncertain.

Our data for the Lorentz number which measures magnetic field strength are fully compatible with a cubic root dependence on the non-dimensional power (after correcting for that part of the energy flux that does not contribute to magnetic field generation). This makes the scaling independent of the rotation rate, which is a central point in the scaling theory of \citet{Davidson2013}. While both dipolar and multipolar dynamos follow basically the same scaling law, we find that in the dipolar case the field strength is larger by $\approx30\%$. In \citet{Yadav2013} we attributed this difference to a slightly weaker flow velocity in dipolar dynamos, but in our present study we do not find a significant difference in the Rossby number between dipolar and multipolar cases. However, we note that the ohmic dissipation time is systematically shorter for multipolar dynamos (open symbols in Fig.~\ref{tmag}) than it is for dipolar dynamos (filled symbols), i.e., the generated magnetic energy is more rapidly dissipated in the multipolar case, leading to a lower mean field strength. The reason for the longer dissipation time of dipolar dynamos could be that more magnetic energy is stored in large spatial scales where it is less vulnerable to dissipative destruction.

Although we find that Boussinesq and anelastic dynamos are described by the same scaling laws, it must be noted that qualitative differences exist between the two. As compared to the Boussinesq models, flow velocity in anelastic models is faster and has smaller length scales in the outer parts of the fluid shell as compared to the deeper interior~\citep{Browning2008, Gastine2012}. This difference affects the nature of the resulting dynamo. For instance, \citet{Gastine2012dyn} find that dipolar dynamos give away to multipolar ones as the density stratification is increased (keeping other parameters the same). They argue that separation of the flow length scales makes the flow structure in highly stratified cases similar to convection in thin shell geometries which prefer non-dipolar dynamos~\citep{Stanley2005}. However, \citet{Duarte2013} show that models with variable electrical conductivity can push the dynamo generation to the deep interior and dipolar dynamos can be easily obtained for even highly stratified cases.

The mismatch between the control parameters of numerical simulations and the astrophysical objects is mainly caused by the expected extremely low values of the various transport coefficients in the latter. The scaling laws we present here are expressed in terms of non-dimensional buoyancy power per unit mass, $P$, which does not involve any diffusive parameters (barring the small $P_{m}$ dependence). Our simulations span $10^{-8}\lesssim P\lesssim 10^{-2}$. To put this range into perspective, we calculate $P$ for the rapidly rotating M4.5 star {\it EV Lac} using the parameters quoted in \citet[Supplementary information]{Christensen2009}. The resulting value, $P\approx 2\times 10^{-8}$, lies within the range covered by our database. As shown by \citet{Christensen2009} the observed surface field strength roughly agrees with the power-based scaling law.

There are few outstanding issues which need further exploration. As has recently been emphasized by \citet{Stelzer2013}, some dependence on the magnetic Prandtl number is required for an optimal fit of the dynamo simulation data. Although the influence of $P_m$ in the scaling laws is secondary, its value in natural dynamos is several orders of magnitude lower than it is in the simulations. Ignoring or including $P_m$ in the scaling laws makes a difference of nearly an order of magnitude in the prediction for planets or stars. \citet{Christensen2004} argued that the $P_m$-dependence, although needed for fitting the model data at $P_m \approx 1$, would disappear in the astrophysically relevant range $P_m \ll 1$ and \citet{Christensen2006} found good agreement between prediction and observation in case of the Earth's dynamo only when the dependence on the magnetic Prandtl number is ignored. However, the assumption that the $P_m$-dependence vanishes at small values has not yet been confirmed by theory or by numerical simulations, which are not available for $P_m$ much smaller than one.

Another issue is the independence of magnetic field scaling from rotation rate that is implied by our scaling law. For low-mass stars a saturated (rotation-rate independent) magnetic surface flux is observed for very rapid rotators with low Rossby number, but for larger values of the Rossby number it decreases roughly proportionally with decreasing rotation rate~\citep[e.g.][]{Reiners2009}. \citet{Schrinner2013} argues that the rotational dependence at large Rossby number may enter through its influence on the ohmic dissipation factor $f_{ohm}$, challenging the often made assumption that $f_{ohm} \approx 1$ in all kinds of natural dynamos. 

One problem with applying our scaling laws to stars and planets is that they predict average values inside the dynamo, whereas observations usually relate to the strength of the magnetic field or part of the field at the surface of the object. The assumption that the two values can be related by a fixed factor is probably too simplistic. It is conceivable that a dependence of this factor on the rotation rate could contribute to the observed dependence of the surface magnetic flux on the Rossby number in M-stars with moderate to low rotation rates.

Our work extends earlier studies~\citep{Christensen2006, Takahashi2008, Aubert2009, Christensen2010, Schrinner2012, Yadav2013}. It shows that previously found scaling laws for velocity and magnetic field strength in rotating dynamos also apply when strong radial variations in density or transport properties are present, as expected in the conducting cores of gas planets, brown dwarfs and low-mass stars. This strongly supports the notion that the power-based scaling laws are rather universal. For the magnetic field scaling there is a decent agreement with observation for several solar system planets and certain classes of stars~\citep{Christensen2009, Christensen2010}. However, the velocity scaling has not yet been tested against observation (except for Earth's core). This together with the open questions on the  role of the magnetic Prandtl number, the cause for weaker surface magnetic fields at less rapidly rotating stars, and the relation between internal and surface magnetic fields  keep this research area vibrant and provide an exciting avenue for future explorations.

\acknowledgements
We thank the anonymous referee for helping to improve the presentation of our paper. We acknowledge funding from the Deutsche Forschungsgemeinschaft (DFG) through Project SFB 963 / A17 and through the special priority program 1488 (PlanetMag, \url{http://www.planetmag.de}) of the DFG. Simulations were run on the GWDG computational facility, G{\"o}ttingen and the Regionales Rechenzentrum f{\"u}r Niedersachsen (RRZN) at Leibniz University Hannover operating under the Norddeutscher Verbund zur F{\"o}rderung des Hoch- und H{\"o}chstleistungsrechnens (HLRN). All the figures were generated using {\it matplotlib} (\url{www.matplotlib.org}).


\begin{thebibliography}{45}
\expandafter\ifx\csname natexlab\endcsname\relax\def\natexlab#1{#1}\fi

\bibitem[{Aubert(2005)}]{Aubert2005}
Aubert, J. 2005, J. Fluid Mech., 542, 53

\bibitem[{Aubert {et~al.}(2001)Aubert, Brito, Nataf, Cardin, \&
  Masson}]{Aubert2001}
Aubert, J., Brito, D., Nataf, H.-C., Cardin, P., \& Masson, J.-P. 2001, Phys.
  Earth Planet. Int., 128, 51

\bibitem[{Aubert {et~al.}(2009)Aubert, Labrosse, \& Poitou}]{Aubert2009}
Aubert, J., Labrosse, S., \& Poitou, C. 2009, Geophys. J. Int., 179, 1414

\bibitem[{{B{\"o}hm-Vitense}(1958)}]{Bohm1958}
{B{\"o}hm-Vitense}, E. 1958, \zap, 46, 108

\bibitem[{Braginsky \& Roberts(1995)}]{Braginsky1995}
Braginsky, S.~I. \& Roberts, P.~H. 1995, Geophys. Astrophys. Fluid Dyn., 79, 1

\bibitem[{Browning(2008)}]{Browning2008}
Browning, M.~K. 2008, ApJ, 676, 1262

\bibitem[{Brun {et~al.}(2004)Brun, Miesch, \& Toomre}]{Brun2004}
Brun, A.~S., Miesch, M.~S., \& Toomre, J. 2004, ApJ, 614, 1073

\bibitem[{Christensen \& Wicht(2007)}]{Christensen2007}
Christensen, U. \& Wicht, J. 2007, in Treatise of Geophysics, Vol.~8, ed.
  G.~Schubert (Amsterdam: Elsevier)

\bibitem[{Christensen(2002)}]{Christensen2002}
Christensen, U.~R. 2002, J. Fluid Mech., 470, 115

\bibitem[{Christensen(2010)}]{Christensen2010}
---. 2010, Space Sci. Rev., 152, 565

\bibitem[{Christensen \& Aubert(2006)}]{Christensen2006}
Christensen, U.~R. \& Aubert, J. 2006, Geophys. J. Int., 166, 97

\bibitem[{Christensen {et~al.}(2009)Christensen, Holzwarth, \&
  Reiners}]{Christensen2009}
Christensen, U.~R., Holzwarth, V., \& Reiners, A. 2009, Nature, 457, 167

\bibitem[{Christensen \& Tilgner(2004)}]{Christensen2004}
Christensen, U.~R. \& Tilgner, A. 2004, Nature, 429, 169

\bibitem[{Curtis \& Ness(1986)}]{Curtis1986}
Curtis, S.~A. \& Ness, N.~F. 1986, J. Geophys. Res., 91, 11003

\bibitem[{Davidson(2013)}]{Davidson2013}
Davidson, P. 2013, arXiv preprint arXiv:1302.7140

\bibitem[{Duarte {et~al.}(2013)Duarte, Gastine, \& Wicht}]{Duarte2013}
Duarte, L.~D., Gastine, T., \& Wicht, J. 2013, arXiv preprint arXiv:1210.3245

\bibitem[{French {et~al.}(2012)French, Becker, Lorenzen, Nettelmann,
  Bethkenhagen, Wicht, \& Redmer}]{French2012}
French, M., Becker, A., Lorenzen, W., Nettelmann, N., Bethkenhagen, M., Wicht,
  J., \& Redmer, R. 2012, ApJS, 202, 5

\bibitem[{Gastine {et~al.}(2012)Gastine, Duarte, \& Wicht}]{Gastine2012dyn}
Gastine, T., Duarte, L., \& Wicht, J. 2012, A\&A, 546, A19

\bibitem[{Gastine {et~al.}(2013)Gastine, Morin, Duarte, Reiners, Christensen,
  \& Wicht}]{Gastine2013a}
Gastine, T., Morin, J., Duarte, L., Reiners, A., Christensen, U., \& Wicht, J.
  2013, A\&A, 549, L5

\bibitem[{Gastine \& Wicht(2012)}]{Gastine2012}
Gastine, T. \& Wicht, J. 2012, Icarus, 219, 428

\bibitem[{Gilman \& Glatzmaier(1981)}]{Gilman1981}
Gilman, P.~A. \& Glatzmaier, G.~A. 1981, ApJS, 45, 335

\bibitem[{Glatzmaier(1984)}]{Glatzmaier1984}
Glatzmaier, G.~A. 1984, J. Comp. Phys., 55, 461

\bibitem[{G{\'o}mez-P{\'e}rez {et~al.}(2010)G{\'o}mez-P{\'e}rez, Heimpel, \&
  Wicht}]{Gomez2010}
G{\'o}mez-P{\'e}rez, N., Heimpel, M., \& Wicht, J. 2010, Phys. Earth Planet.
  Int., 181, 42

\bibitem[{Jones {et~al.}(2011)Jones, Boronski, Brun, Glatzmaier, Gastine,
  Miesch, \& Wicht}]{Jones2011b}
Jones, C., Boronski, P., Brun, A., Glatzmaier, G., Gastine, T., Miesch, M., \&
  Wicht, J. 2011, Icarus, 216, 120

\bibitem[{Jones(2011)}]{Jones2011a}
Jones, C.~A. 2011, Annual Rev. of Fluid Mech., 43, 583

\bibitem[{Jones {et~al.}(2009)Jones, Kuzanyan, \& Mitchell}]{Jones2009b}
Jones, C.~A., Kuzanyan, K.~M., \& Mitchell, R.~H. 2009, J. Fluid Mech., 634,
  291

\bibitem[{Kaspi {et~al.}(2009)Kaspi, Flierl, \& Showman}]{Kaspi2009}
Kaspi, Y., Flierl, G.~R., \& Showman, A.~P. 2009, Icarus, 202, 525

\bibitem[{Kunnen {et~al.}(2010)Kunnen, Geurts, \& Clercx}]{Kunnen2010}
Kunnen, R., Geurts, B., \& Clercx, H. 2010, J. Fluid Mech., 642, 445

\bibitem[{Lantz \& Fan(1999)}]{Lantz1999}
Lantz, S. \& Fan, Y. 1999, ApJS, 121, 247

\bibitem[{Mizutani {et~al.}(1992)Mizutani, Yamamoto, \&
  Fujimura}]{Mizutani1992}
Mizutani, H., Yamamoto, T., \& Fujimura, A. 1992, Adv. Space Res., 12, 265

\bibitem[{Olson \& Christensen(2006)}]{Olson2006}
Olson, P. \& Christensen, U.~R. 2006, Earth and Planet. Sci. Lett., 250, 561

\bibitem[{Ossendrijver(2003)}]{Ossendrijver2003}
Ossendrijver, M. 2003, A\&A Rev., 11, 287

\bibitem[{Reiners {et~al.}(2009)Reiners, Basri, \& Browning}]{Reiners2009}
Reiners, A., Basri, G., \& Browning, M. 2009, \apj, 692, 538

\bibitem[{Sano(1993)}]{Sano1993}
Sano, Y. 1993, J. Geomagn. Geoelectr., 45, 65

\bibitem[{Schrinner(2013)}]{Schrinner2013}
Schrinner, M. 2013, MNRAS Let., 431, L78

\bibitem[{Schrinner {et~al.}(2012)Schrinner, Petitdemange, \&
  Dormy}]{Schrinner2012}
Schrinner, M., Petitdemange, L., \& Dormy, E. 2012, ApJ, 752, 121

\bibitem[{Showman {et~al.}(2011)Showman, Kaspi, \& Flierl}]{Showman2011}
Showman, A.~P., Kaspi, Y., \& Flierl, G.~R. 2011, Icarus, 211, 1258

\bibitem[{Stanley {et~al.}(2005)Stanley, Bloxham, Hutchison, \&
  Zuber}]{Stanley2005}
Stanley, S., Bloxham, J., Hutchison, W.~E., \& Zuber, M.~T. 2005, Earth and
  Planet. Sci. Lett., 234, 27

\bibitem[{Starchenko \& Jones(2002)}]{Starchenko2002}
Starchenko, S. \& Jones, C. 2002, Icarus, 157, 426

\bibitem[{Stelzer \& Jackson(2013)}]{Stelzer2013}
Stelzer, Z. \& Jackson, A. 2013, Geophys. J. Int., 193, 1265

\bibitem[{Stevenson(1979)}]{Stevenson1979}
Stevenson, D.~J. 1979, Geophys. Astrophys. Fluid Dyn., 12, 139

\bibitem[{Takahashi {et~al.}(2008)Takahashi, Matsushima, \&
  Honkura}]{Takahashi2008}
Takahashi, F., Matsushima, M., \& Honkura, Y. 2008, Phys. Earth Planet. Int.,
  167, 168

\bibitem[{Wicht(2002)}]{Wicht2002}
Wicht, J. 2002, Phys. Earth Planet. Int., 132, 281

\bibitem[{Wicht \& Tilgner(2010)}]{Wicht2010}
Wicht, J. \& Tilgner, A. 2010, Space Sci. Rev., 152, 501

\bibitem[{Yadav {et~al.}(2013)Yadav, Gastine, \& Christensen}]{Yadav2013}
Yadav, R.~K., Gastine, T., \& Christensen, U.~R. 2013, Icarus, 225, 185

\end{thebibliography}

\end{document}